\documentclass[doublecol]{epl2}
\usepackage{float}
\usepackage{amssymb}
\usepackage{amsmath}
\usepackage[figtopcap]{subfigure}
\usepackage{bm}
\usepackage{color}
\usepackage{graphics}
\usepackage{color}

\title{Complete hyperfine Paschen-Back regime at relatively small magnetic fields realized in Potassium nano-cell}
\shorttitle{Potassium nano-cell} 

\author{A. Sargsyan\inst{1} \and A. Tonoyan\inst{1,2} \and G. Hakhumyan\inst{1}
\and C. Leroy\inst{2} \and Y. Pashayan-Leroy\inst{2} \and D. Sarkisyan\inst{1} }
\shortauthor{A. Sargsyan \etal}
\institute{
  \inst{1} Institute for Physical Research, NAS of Armenia  - Ashtarak-2, 0203, Armenia\\
  \inst{2} Laboratoire Interdisciplinaire Carnot de Bourgogne, UMR CNRS 6303, Universit\'{e} de Bourgogne  - Dijon, France
}


\abstract{
A one-dimensional nano-metric-thin cell (NC) filled with potassium metal has been built and used to
study optical atomic transitions in external magnetic fields. These studies benefit from the remarkable features
of the NC allowing one to use $\lambda$/2- and $\lambda$-methods for effective investigations of individual
transitions of the K $D_1$ line. The methods are based on strong narrowing of the absorption spectrum of the atomic
column  of thickness $L$ equal to $\lambda$/2 and to $\lambda$ (with $\lambda = 770\un{nm}$  being the resonant
laser radiation wavelength). In particular, for a  $\pi$-polarized radiation excitation the $\lambda$-method
allows us to resolve eight atomic transitions (in two groups of four atomic transitions) and
to reveal two remarkable transitions that we call Guiding Transitions (GT). The probabilities of all
other transitions inside the group (as well as the frequency slope versus magnetic field) tend to
the probability and to the slope of GT. Note that for circular polarization there is
one group of four transitions and GT do not exist. Among eight transitions there are also
two transitions (forbidden for $B$ = 0) with the probabilities undergoing strong modification
under the influence of magnetic fields. Practically the complete hyperfine Paschen-Back regime
is observed at relatively low ($\sim 1\un{kG}$) magnetic fields. Note that for K $D_2$ line GT are absent.
Theoretical models describe the experiment very well. }
\begin{document}

\maketitle
\section{Introduction}
Atomic spectroscopy with NC filled with Rb or Cs atomic vapor, with a thickness of the vapor
column $L$ which is of the order of optical radiation wavelength   has been found to be very efficient to
study optical atomic transitions in external magnetic fields~\cite{Sarg_2008}. There are two interconnected effects:
splitting of the atomic energy levels to Zeeman sublevels and shifting of frequencies (deviating from
the linear dependence observed in quite moderate magnetic fields), and significant change in atomic transitions
probabilities as a function of the $B$ field~\cite{Tremblay,Bud1,Auz1}. These studies benefit from the following features of NC:
i) sub-Doppler spectral resolution for atomic vapor thickness $L = \lambda/2$ and $L = \lambda$ ($\lambda$ being the resonant
wavelength of Rb $D_{1,2}$ or Cs $D_{1,2}$ lines) needed to resolve a huge number of Zeeman transition
components in transmission or fluorescence spectra; ii) possibility to apply a strong magnetic field
using permanent magnets in spite of a strong inhomogeneity of the $B$ field (in our case it can reach $150\un{G/mm}$).
Note that the variation of the $B$ field inside the atomic vapor is negligible as the vapor column thickness is small.
K vapor in magnetic fields was studied in~\cite{Bloch1} using a few centimeter-long cell and saturation absorption technique.
 However, due to the presence of strong cross-over resonances in the absorption spectrum, the technique is useful only
 for $B<10\un{G}$.\\
 \indent Here we report the first studies of $K$ vapor confined in NC and under the influence of
 relatively low magnetic fields ($B<2\un{kG}$). The NC with $L= \lambda/2$ and $\lambda$ are used for $\sigma^+$ and  $\pi$ laser
 polarizations. Also two theoretical considerations are taken into account. The splitting of atomic
 levels in weak magnetic fields is described by the total angular momentum $\textbf{F = J + I}$ of the atom
 and its projection $m_F$, where $\textbf{J = L + S}$ is the total angular momentum of electrons and $\textbf{I}$ is the
 nuclear spin. In the hyperfine Paschen-Back (HPB) regime, $\textbf{J}$ and $\textbf{I}$ become decoupled and the
 splitting of the atomic levels is described by the projections $m_J$ and $m_I$. For alkali metals
 the hyperfine Paschen-Back regime takes place at fields $B \gg B_0 = A_{hfs}/\mu_B$, where $ A_{hfs}$ is the
 ground-state hyperfine coupling coefficient and $\mu_B$ is the Bohr magneton. For $^{133}$Cs, $^{87}$Rb,
 and $^{85}$Rb atoms $B_0 \sim 1.7\un{kG}, 2\un{kG}$ and $0.7\un{kG}$, respectively~\cite{Musso,Happer,Sarg1,Weller,Sarg2}. It is worth  noting
 that $^{39}$K  has the smallest $B_0  \approx 0.17\un{kG}$ caused by the smallest hyperfine splitting $\Delta \approx 462\un{MHz}$
 of the ground level $4S_{1/2}$ [note that $A_{hfs} = \Delta / (I+1/2)$]. Thus, one expects to obtain
 complete $J$ and $I$ decoupling (HPB) of $^{39}$K at relatively low magnetic fields $B \gg 170\un{G}$.
 The manifestations of HPB regime, particularly, are as follows: i) strong reduction of
 the number of atomic transitions (compared with that at low magnetic field) to a fixed
 number which is easy to determine from the diagram on the basis of $m_J$ and $m_I$ projections,
 ii) the fixed frequency slope inside the group of transitions, iii) the transition
 probabilities tend to the same value inside the group, iv) the energy of the ground
 and upper levels can be calculated from analytic expression~(\ref{eq:E}).
 \begin{figure}
\onefigure[scale=0.16] {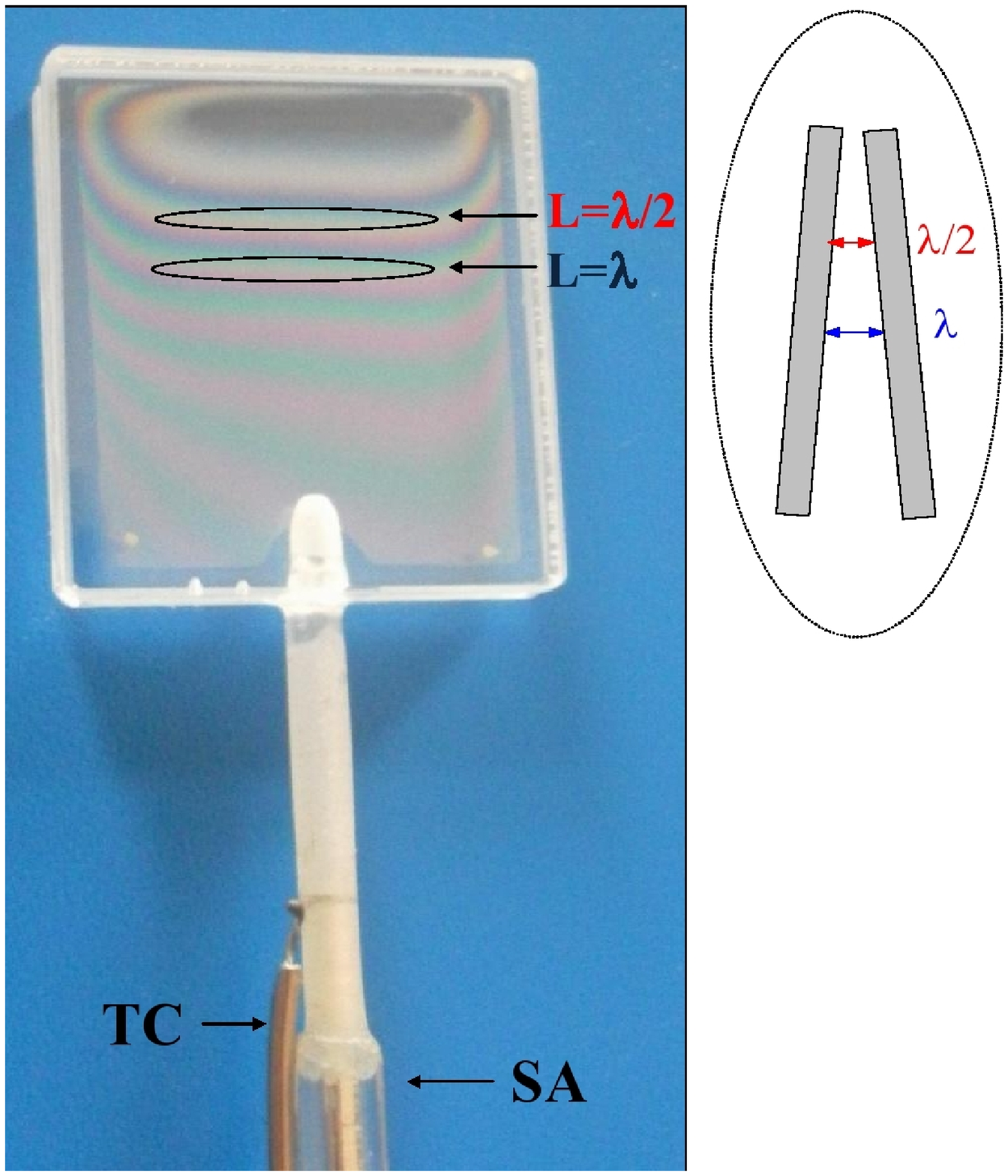}
\caption{The photograph of the K NC with a tapered gap of thickness $L$ ranging from $50$ to $1500\un{nm}$.
TC is a thermocouple to measure the side-arm (SA) temperature.}
\label{fig:1}
\end{figure}
\begin{figure}
\onefigure[scale=0.4] {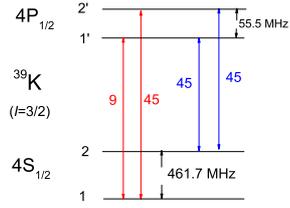}
\caption{The hyperfine (hfs) energy levels diagram of the $D_1$ line of $^{39}$K
(the prime is used for the upper levels).}
\label{fig:2}
\end{figure}
 \section{Experiment}
  \begin{figure}[h!]
\onefigure[scale=0.5] {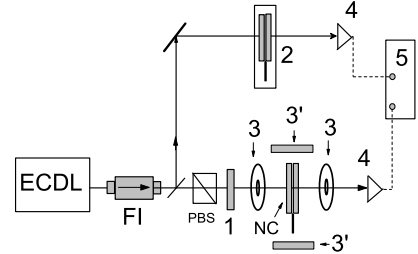}
\caption{Sketch of the experimental setup. ECDL, diode laser; FI, Faraday isolator; 1, $\lambda/4$ plate; main K
NC with thickness $L = \lambda/2$ or $L = \lambda$ in the oven; PBS, polarizing beam splitter; 2, auxiliary K NC;
$3,3^{\prime}$
permanent magnets: in case of $\sigma^{+}$ radiation use, the $\textbf{B}$-field is directed along the laser propagation
direction $\textbf{k}$ (magnets 3); for $\pi$ polarization, the $\textbf{B}$-field is directed along the laser electric field $\textbf{E}$
(magnets $3^{\prime}$); 4, photodetectors; 5, oscilloscope.}
\label{fig:3}
\end{figure}
\begin{figure}[h!]
\subfigure[]{
\resizebox{0.2\textwidth}{!}{\includegraphics{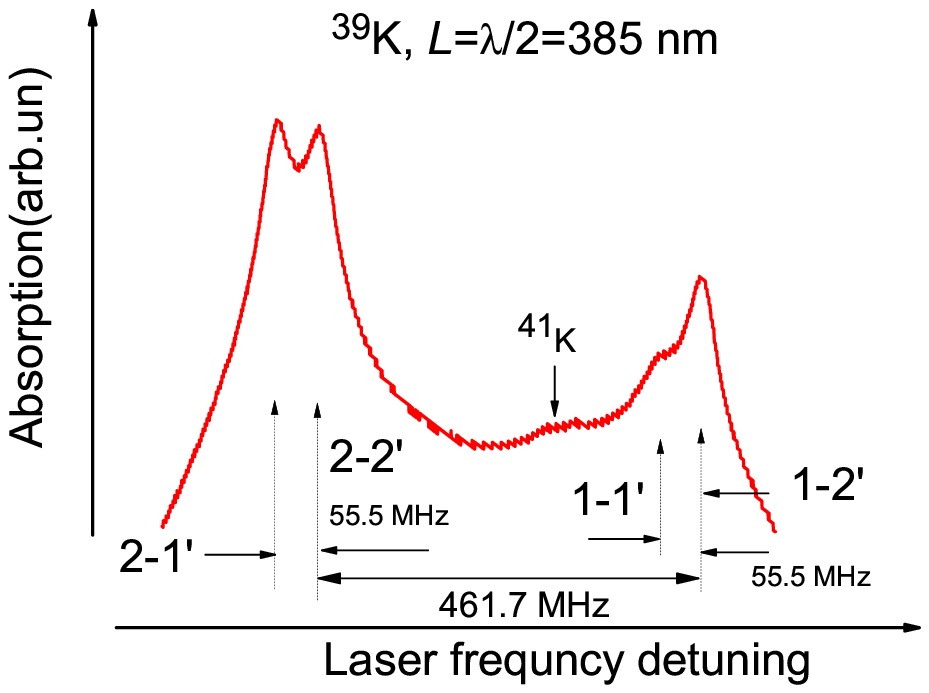}
}
\label{fig:4a}
} \hspace{0.005cm}
\subfigure[]{
\resizebox{0.2\textwidth}{!}{\includegraphics{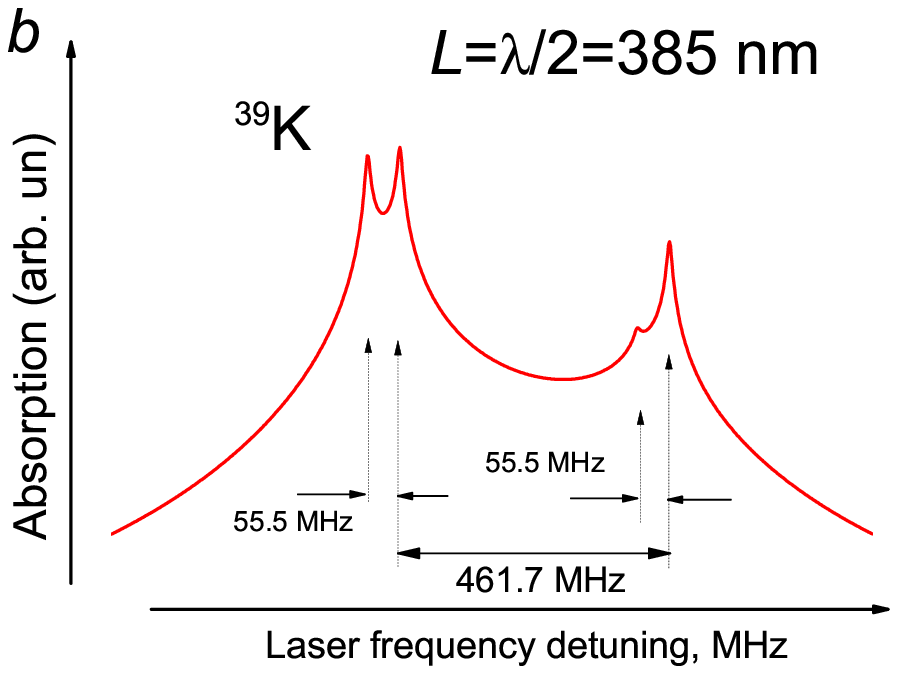}
}
\label{fig:4b}
}
\caption{Absorption spectrum of $^{39}$K vapor contained in the NC $L=\lambda/2=385$~nm a) experiment, the laser power is 2~$\mu$W,
the SA temperature  is $150\un{^{\circ}C}$, a small absorption of $^{41}$K is also seen; b) theory [16] with  the parameters:
thermal velocity $v_{th}=450\un{ms^{-1}}$, Rabi frequency  $\Omega/2\pi = 0.06 \gamma_N$ ($\gamma_N \approx  6$~MHz).}
\label{fig:4}
\end{figure}

 \begin{figure}[hbtp]
\onefigure[scale=0.65] {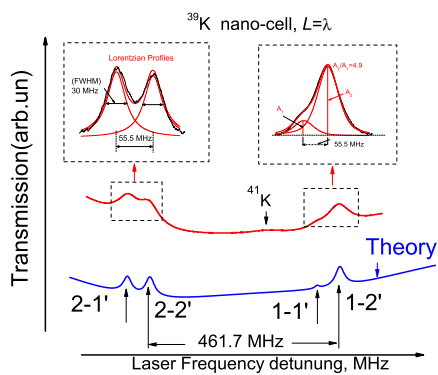}
\caption{Four VSOPs located at the positions of the transitions $1,2 \rightarrow 1^{\prime},2^{\prime}$ are presented
in the transmission spectrum of $^{39}$K vapor contained in the NC with $L =770\un{nm}$; a small
absorption of $^{41}$K is also seen; the SA temperature is $160\un{^{\circ}C}$. The inset shows the results of fitting
by Lorentzian profiles; note that the ratio of the VSOPs amplitudes coincides
with that deduced from the values given in Fig.~\ref{fig:2} which means linear dependence on laser power.
The lower curve - the theory~\cite{JOSAB_2007} for the parameters: $v_\mathrm{{th}}=450\un{m s^{-1}}$, Rabi
frequency $\Omega/2\pi = 0.2 \gamma_N$.}
\label{fig:5}
\end{figure}
\begin{figure}[hbtp]
\onefigure[scale=0.65] {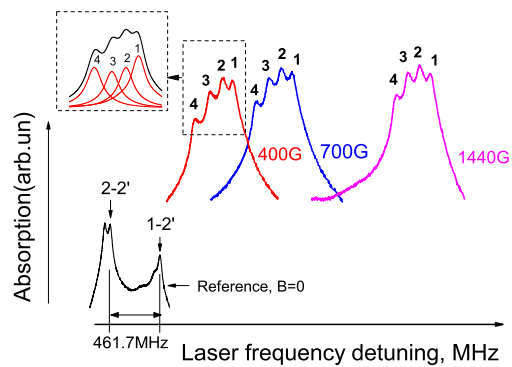}
\caption{Absorption spectrum of $^{39}$K vapor contained in the NC with $L = \lambda/2$ for $B=400,
700$ and $1440\un{G}$;  $\sigma^{+}$ excitation. The bottom curve is the reference showing the
positions of $^{39}$K transitions for $B=0$. The inset shows the
results of fitting by four "pseudo-Vigt" functions (note that the amplitude of the
transition labeled $\textit{4}$ is larger than that of $\textit{3}$). The absolute value
of the peak absorption of transition $\textit{1}$ is $\sim 0.3\%$.}
\label{fig:6}
\end{figure}

\begin{figure}[hbtp]
\onefigure[scale=0.45] {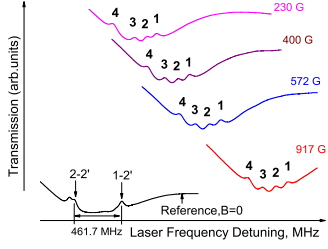}
\caption{Transmission spectrum of $^{39}$K vapor contained in the NC with $L = \lambda$  for $B=230, 400,572$ and 917~G
and $\sigma^+$ excitation. The bottom curve is the reference showing the
positions of $^{39}$K transitions for $B=0$. }
\label{fig:7}
\end{figure}

\begin{figure}[hbtp]
\onefigure[scale=0.4] {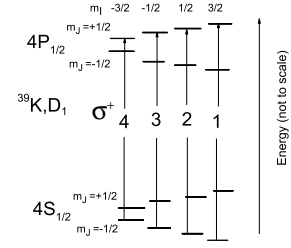}
\caption{Diagram of $^{39}$K $D_1$ line transitions in HPB regime for $\sigma^{+}$ laser excitation.
The selection rules: $\Delta m_J=1$; $\Delta m_I=0$.}
\label{fig:8}
\end{figure}
\begin{figure}[hbtp]
\onefigure[scale=0.5] {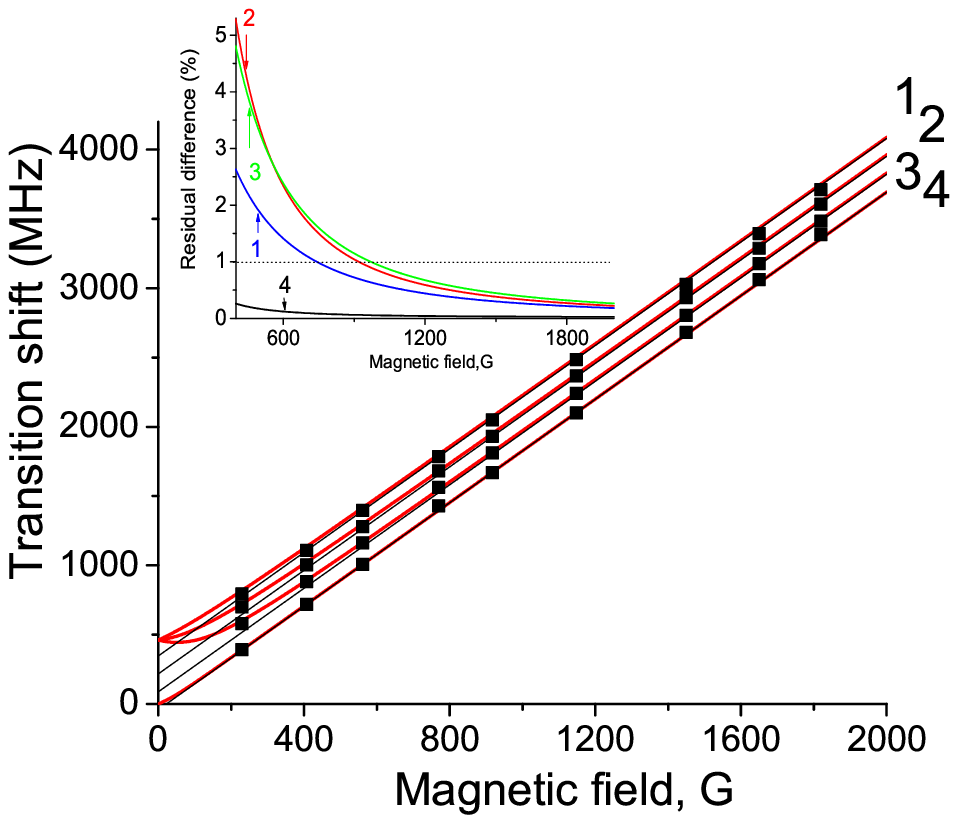}
\caption{$B$-field dependence of the frequency shifts for transitions $\textit{1- 4}$ of the $^{39}$K $D_1$
line in case of $\sigma^+$ excitation. Red lines are given by exact numerical calculations,
black solid lines by formula~(\ref{eq:E}), black squares by experiment.
The inaccuracy is $\sim2\%$. Inset  presents the residual difference
between numerics and calculations by formula (1) for curves $\textit{1-4}$.}
\label{fig:9}
\end{figure}
 For our purpose a one-dimensional nano-metric-thin cell filled with  natural potassium (93.3\% $^{39}$K, and 6.7\% $^{41}$K)
 has been built (for the first time) and used for the experiment. The design of the NC is similar
 to that of the extremely thin cell described earlier~\cite{Sark2} (the details of NC design can be found in~\cite{Sarg3}).
  The NC allows one to exploit a variable vapor column thickness $L$ in the range of $50-1500\un{nm}$.
  It is demonstrated that for K vapor the key parameter
   determining the spectral width and the shape of the absorption line in the NC is the
   ratio $L/\lambda$, with $\lambda=770\un{nm}$ being the wavelength of the laser radiation resonant with the atomic
   transition of $D_1$ line (as it was earlier demonstrated for the NC filled with Cs or Rb). In particular, it
   was shown that the spectral width of the resonant absorption reaches its
   minimum value at $L = (2n +1) \lambda/2$ ($n$ is an integer); this effect has been called the
   Dicke-type coherent narrowing~\cite{Ducl,Sarg3,Sark3}. It is also demonstrated that for $L = n\lambda$ the
   spectral width of the resonant absorption reaches its maximum value, close to the Doppler width ($>0.9\un{GHz}$).\\
   \indent A photograph of the K NC with a tapered gap is shown in Fig.~\ref{fig:1}.
   The windows of the NC were constructed with well-polished crystalline sapphire with the $c$
    axis perpendicular to the window surface to minimize birefringence. The regions with the thickness
    $L = \lambda/2 = 385\un{nm}$ and $L = \lambda = 770\un{nm}$ are marked by ovals in Fig.~\ref{fig:1}. A sapphire side-arm (SA)
    filled with metallic K is seen at the bottom of Fig.~\ref{fig:1}. The SA was heated to $150-160\un{^{\circ}C}$ (the
    temperature on the windows was by $20$ degrees higher in order to prevent vapor condensation) providing
    the density of atoms $N = (5-8) \times 10^{12}\un{cm}^{-3}$. The NC can operate up to temperatures of
    $500\un{^{\circ}C}$. The hfs energy levels diagram of the $D_1$ line of $^{39}$K is shown in Fig.~\ref{fig:2}. A sketch of the experimental setup is shown in Fig.~\ref{fig:3}. The linearly polarized beam
    of an extended cavity diode laser ($\gamma_L < 1\un{MHz}$), resonant with a $^{39}$K $D_1$ line after passing
    through a Faraday isolator (FI), was focused onto a $0.5\un{mm}$ diameter spot on the K NC orthogonally
    to the cell window. A PBS was used to purify the initial linear polarization of the laser; a $\lambda/4$ plate (1)
    was used to produce a circular polarization. In the experiments the thicknesses of the vapor
    column $L =\lambda$ and $L= \lambda/2$ were used. The transmission signal was detected
    by a photodiode $(4)$ and was recorded by a Tektronix TDS 2014B four-channel storage oscilloscope $(5)$.
    To record the transmission spectra, the laser radiation was linearly scanned within up to a $\sim 5\un{GHz}$ spectral
    region covering the studied group of transitions. About 30\% of the laser power was branched to
    the reference unit with an auxiliary K NC (2). The absorption spectrum of the latter
    with thicknesses $L=\lambda/2$ or $L=\lambda$ was used as frequency reference.\\
    \indent The experimental and theoretical absorption spectra of K atomic vapor contained in
    the NC with $L= ~\lambda/2=385\un{nm}$ are shown in Fig.~\ref{fig:4} (the large width of the spectrum at the base is caused by
    huge Doppler-broadening $> 0.9\un{GHz}$ at $170\un{^{\circ}C}$ at the NC windows). There is a strong spectral
    narrowing which allows one to separate four transitions between hyperfine sublevels shown
    in Fig.~\ref{fig:2}. The sharp (nearly Gaussian) absorption near the top makes it convenient to
    separate closely spaced individual atomic transitions in an external magnetic field
    (we called it  $\lambda/2$-method~\cite{Sarg2}). We have also used the so called  $\lambda$-method~\cite{Sark3,JOSAB_2007}.
    In this case spectrally narrow velocity selective optical pumping (VSOP) resonances located exactly
    at the positions of atomic transitions appear in the transmission spectrum of the NC with thickness
    $L=\lambda$  shown in Fig.~\ref{fig:5}. The VSOP parameters are shown to be immune against 10\% thickness deviation
    from $L= \lambda$, which makes the $\lambda$-method feasible. Laser power is $\sim 15\un{\mu W}$(for a lower power the VSOP
    line-width is less than $30\un{MHz}$, but is more noisy). In a magnetic field, the VSOPs are split
    into many components. The amplitude and frequency
    positions of VSOPs depend on the $B$-field, which makes it convenient to study each individual transition separately~\cite{Sarg_2008}.
 \begin{figure}[hbtp]
\onefigure[scale=0.45] {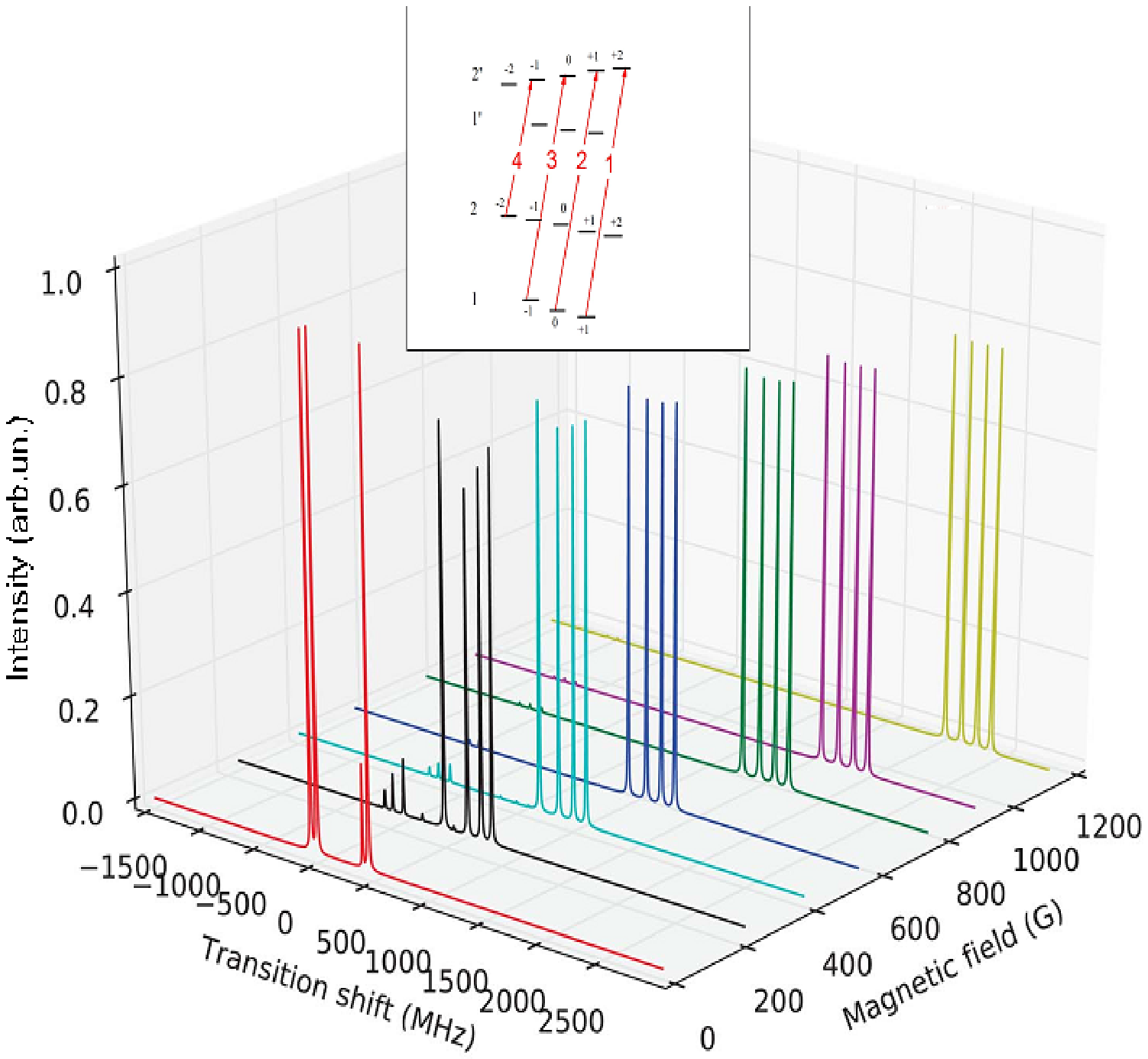}
\caption{The intensities and frequencies shifts (theory) of the $^{39}$K transitions versus $B$-field
for $\sigma+$ excitation. As seen for $B>B_0$ only $\textit{1-4}$ transitions remain and the probabilities tend
asymptotically to the same value at $B \gg B_0$. The inset shows $F, m_F$ for the ground and the upper levels for
transitions $\textit{1-4}$ .}
\label{fig:10}
\end{figure}
    \section{Magnetometry with  $\lambda/2$- and  $\lambda$-methods, $\sigma^{+}$ laser excitation}
    The assembly of an oven with NC inside was placed between the permanent ring magnets. The magnetic
    field $\textbf{B}$ was directed along the laser radiation propagation direction $\textbf{k}$. An extremely small thickness of
    the NC is advantageous for the application of strong magnetic fields with the use of permanent magnets
    (having a $2\un{mm}$ diameter hole for the laser beam passage). Such magnets are unusable for centimeter-long cells
    because of the strong inhomogeneity of the magnetic field, while in NC, the variation of the $B$-field inside
    the atomic vapor column is by several orders less than the applied $B$ value. The $B$-field strength was
    measured by a calibrated Hall gauge with an absolute imprecision less than $30$~G throughout the applied $B$-field range.
    The absorption and transmission spectra of $^{39}$K vapor contained in the NC with $L = \lambda/2$ and $L = \lambda$
    (for $\sigma^+$ radiation) versus magnetic field are shown in Fig.~\ref{fig:6} and Fig.~\ref{fig:7}, respectively (the bottom curves are the reference ones). Four atomic transitions (as predicted for the HPB regime by the diagram presented in Fig.~\ref{fig:8}) are well observable.
    Note that the frequency separation between the two well resolved VSOP labeled 1 and 2 is about $90\un{MHz}$,
    which is more than 10 times less than the Doppler line-width. Also, the VSOPs  have the same amplitudes which is an additional evidence of HPB regime. Let us briefly stress the benefits of $\lambda/2$- and $\lambda$-methods.
    The $\lambda/2$-method requires less
    laser intensity and in case of small absorption the peak absorption $A$ is proportional to $\sigma NL$,
    where $\sigma$ is the absorption cross section proportional to $d^2$ (with $d$ being the dipole moment).
     Thus, by direct comparison of $A_i$ (peak amplitudes of the absorption of the $i$-th transition),
     it is straightforward to estimate the relative probabilities (i.e. line intensities).
     On the other hand, the $\lambda$-method provides a fivefold better spectral resolution. Thus,
     the methods can be considered as complementary depending on particular requirements.
     Moreover, it is easy to switch from $\lambda/2$ to $\lambda$ just by vertical translation of the NC.
     Note, that in the case where there is a big frequency separation between the transitions
     (as it is for the $^{87}$Rb isotope) also a $1\un{mm}$ long cell can be used~\cite{Weller,Weller2}.\\
     \indent Theoretical simulations for the $B$-field dependences of the atomic transition frequency
     shifts and relative transition probabilities for $1,2 \rightarrow 1^\prime,2^\prime$ transitions of the K $D_1$ line
     were based on the calculation of the eigenvalues and eigenvectors of the Hamilton matrix
     versus magnetic field for the full hyperfine structure manifold~\cite{Tremblay,Auz1,Leroy}.
Although for the K $D_1$ line the Rabi-Breit formula could be also used, however the theoretical model is
preferable since it is valid also for K $D_2$ line.
Note, that in
     the case of HPB regime the energy of the ground $4S_{1/2}$ and upper $4P_{1/2}$ levels for the $^{39}$K $D_1$
     line is given by the following formula~\cite{Tremblay,Auz1}:
     \begin{equation}
\begin{array}{r}
E_{|Jm_{J} Im_{I} \rangle}= A_{hfs} m_J m_I +  \mu_B (g_J m_J + g_I m_I) B.
\end{array}
\label{eq:E}
\end{equation}
     \indent The values for the fine structure ($g_J$) and the nuclear ($g_I$) Land\'{e} factors and the hyperfine
     constants $A_{hfs}$ are given in~\cite{Tiecke}. The dependence of the frequency shifts on the magnetic field
     (relative to the position of the $2 \rightarrow 2^\prime$ transition at $B = 0$) is shown in Fig. 9,
     where the red lines are obtained by exact numerical calculation, while the black lines are plotted
     with formula (1); the black squares are the experimental results. The inaccuracy does not exceed 2\%. A good agreement
     of theory and experiment is observed. The inset shows the residual difference between numerics
     (red lines) and calculations (black lines) with formula~\ref{eq:E}. It is remarkable that at a relatively
     low magnetic field $\sim1000\un{G}$ formula~(\ref{eq:E}) describes the transition frequency values with an
     inaccuracy of $\leq 1\%$  (while at $B=1.8$~kG the difference is $ \leq 0.3\%$ which could be
     considered as practically the full HPB regime).  Fig.~\ref{fig:10} presents the theoretical values of the
     atomic transition probabilities (intensities) as a function of the $B$-field. As seen the intensities
     of  transitions $\textit{1-4}$ at $B \geq 1\un{kG}$ tend to the same asymptotic value (HPB regime), which coincides
     with the experiment, while at $B < 400\un{G}$, the intensity of the transition labeled $4$ (Fig.~\ref{fig:10}) is
      larger than those of $\textit{3}$ and $\textit{2}$ which also coincides with the experiment.
\begin{figure}[hbtp]
\subfigure[]{
\resizebox{0.2\textwidth}{!}{\includegraphics{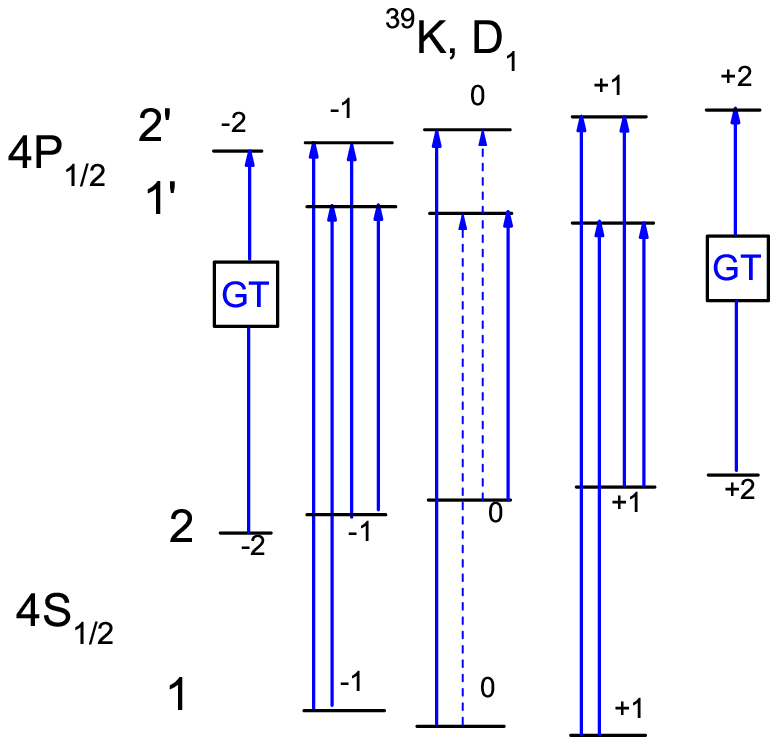}
}
\label{fig:11a}
} \hspace{0.005cm}
\subfigure[]{
\resizebox{0.2\textwidth}{!}{\includegraphics{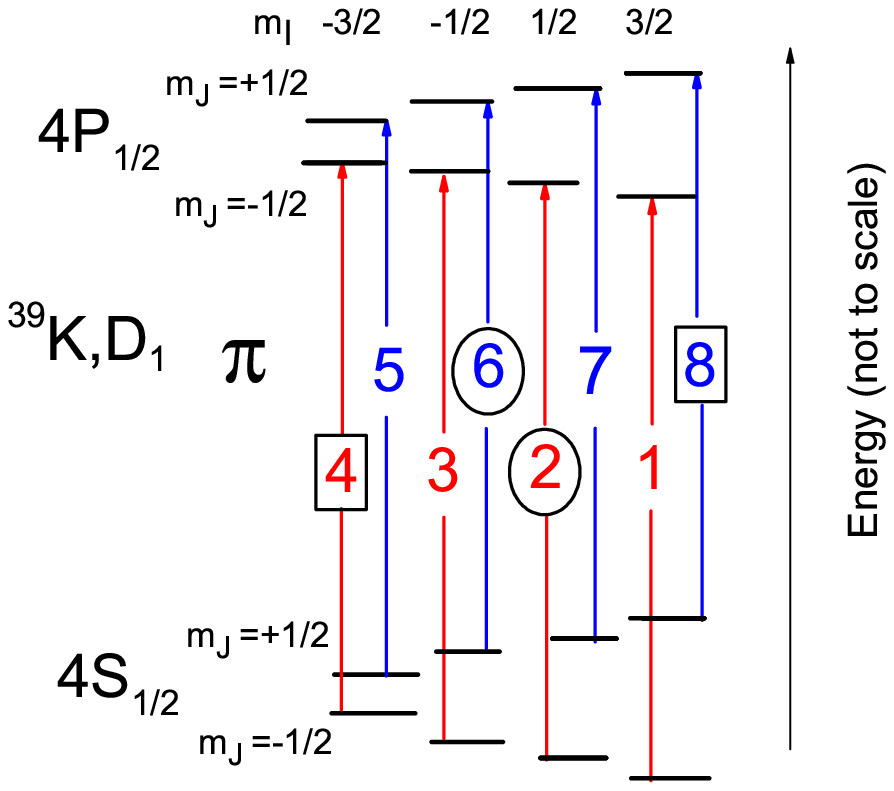}
}
\label{fig:11b}
}
\caption{ a) Diagram of the hfs energy levels of $D_1$ line of $^{39}$K in a magnetic field
(for $B < B_0$) and possible Zeeman transitions for  $\pi$-polarized excitation are shown. The
selection rules are $\Delta F = 0, \pm 1; \Delta m_F= 0$. GT are labeled by rectangles; two IFFA
transitions are indicated by dashed lines; b) Diagram of the K $D_1$ line transitions
in the HPB regime for  $\pi$-polarised excitation. The selection rules are: $\Delta m_J=0; \Delta m_I= 0$.
Two IFFA transitions are labeled by ovals.}
\label{fig:11}
\end{figure}
            \section{Magnetometry with $\lambda/2$- and  $\lambda$-methods,  $\pi$-polarised excitation}
            For this case the $\textbf{B}$-field is directed along the laser electric field $\textbf{E}$
            (magnets $3^{\prime}$ are used and $\lambda/4$ plate is removed).
                  Fig.~\ref{fig:11}\subref{fig:11a}  shows the hfs energy levels diagram of the $D_1$ line of $^{39}$K atoms in a magnetic field ($B < B_0$)
      and 14 possible atomic Zeeman transitions for  $\pi$-polarized excitation [including two forbidden
      at $B=0$ atomic transitions~\cite{Tiecke,Sarg4} shown by dashed lines that we call "initially forbidden
      further allowed" (IFFA)]. Fig.~\ref{fig:11}\subref{fig:11b} presents the diagram of eight remaining transitions
      (the probabilities increase with the $B$-field).\\
\begin{figure}[hbtp]
\onefigure[scale=0.45] {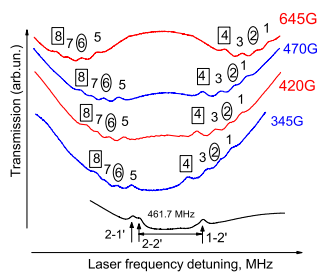}
\caption{Transmission spectrum of the NC with $^{39}$K  for $L = \lambda$ for $B=345, 420, 470$ and $645\un{G}$ ;  $\pi$-polarized
excitation is used. GT and IFFA transitions are labeled by rectangles and ovals, respectively.
The bottom curve is the reference.}
\label{fig:12}
\end{figure}
\begin{figure}[htbp]
\onefigure[scale=0.65] {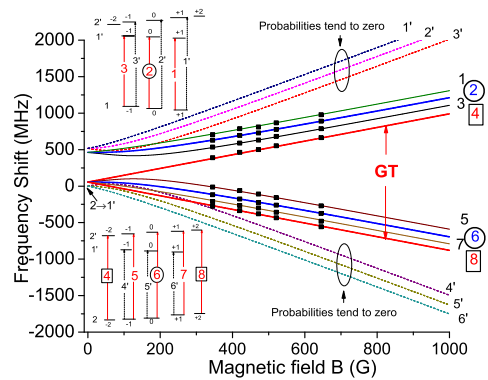}
\caption{$B$-field dependence of the frequency shifts for transitions 1-8 of $^{39}$K $D_1$
line for  $\pi$-polarized excitation. GT and IFFA transitions are marked by rectangles and ovals,
respectively. The probabilities of six transitions (dashed lines) tend to zero for $B>B_0$.
Solid lines - numerics; symbols - experiment (the inaccuracy is $\sim2\%$). The insets show
the initial quantum numbers $F, m_F$ for the ground and upper levels.}
\label{fig:13}
\end{figure}
 \begin{figure}[hbtp]
\onefigure[scale=0.65] {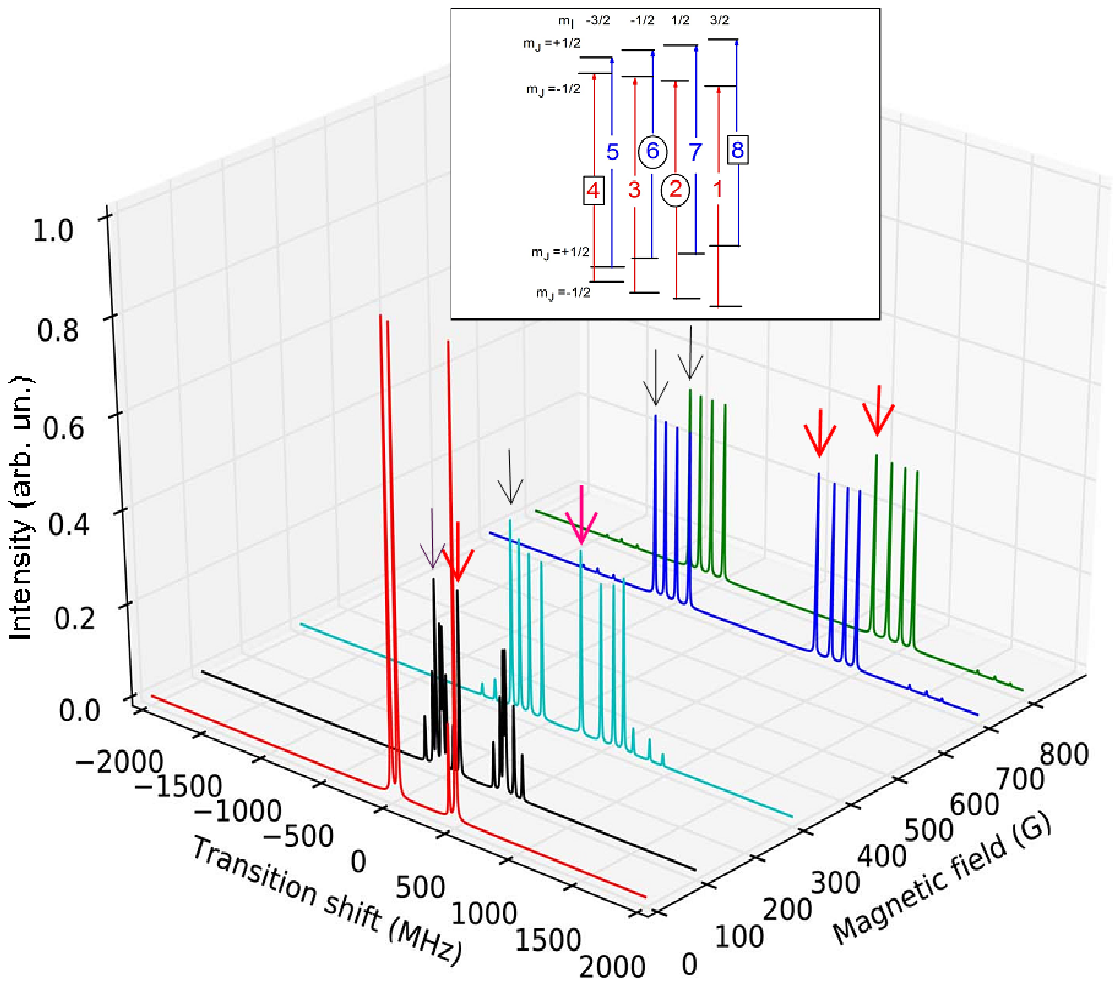}
\caption{Theoretical curves for intensities and frequency shifts of the atomic transitions
of $^{39}$K $D_1$ line vs $B$-field for  $\pi$-polarized  excitation. As seen for $B \gg B_0$ only 1-8 transitions
 remain and the probabilities tend asymptotically to the same value as for GT transitions
  (by module $4.66 \times 10^{-18}\un{ESU}$). Groups 1-4 and 5-8 are located at the high and low frequency
  side, respectively, while the transitions labeled 1 and 8 have the largest and smallest
  frequencies, respectively. GT are marked by arrows (each by the same colour) and as seen their amplitudes remain
  the same in the range of $100 - 800\un{G}$.}
\label{fig:14}
\end{figure}
      \indent Note, that the necessary conditions for a modification of the probability are
      the following~\cite{Leroy,Leroy2}: the perturbation induced by the $B$-field couples only sublevels
      with $m_F -m_F^\prime= 0$ which satisfy the selection rules $\Delta L=0$, $\Delta J=0$, $\Delta F=±1$, and modification of
      the probability is possible only for transitions between ground and excited sublevels when
      at least one of them is coupled with another transition's sublevel according to the selection
      rules. As seen from Fig.~\ref{fig:11}\subref{fig:11a}, for the two side atomic transitions of K (marked as GT),
      $|2, m_F =+2\rangle \rightarrow |2^\prime, m_F{^{\prime}} = +2 \rangle$  and
      $|2, m_F =-2 \rangle \rightarrow |2^\prime, m_F{^{\prime}} = -2 \rangle$ the sublevels which could be mixed
      according to the selection rules are absent. Thus the probability of these two transitions
      remain the same in the whole range of applied $B$-fields, while the probabilities of other
       transitions  differ significantly at low fields, but tend to the same value within the
       group at $B \gg B_0$. It is remarkable that the probabilities of all the other atomic transitions
       tend to that of two side atomic transitions (this is confirmed by the numerical calculations).
       For this reason why we call them guiding transitions. In Fig.~\ref{fig:12} the transmission spectra of  $\pi$-polarized
       excitation of the potassium NC ($L = \lambda$ for $B=345, 420, 470$ and $645\un{G}$) are shown. GT and IFFA
       transitions are labeled by rectangles and ovals, respectively (as seen IFFA transitions undergo strong modification under the influence of the $B$-field).\\
\indent The $B$-field dependence of the frequency shifts for transitions $\textit{1-8}$ for  $\pi$-polarized
       radiation is shown in Fig.~\ref{fig:13}. GT and IFFA transitions are labeled by rectangles and ovals,
       respectively. We see that transitions $\textit{1-8}$  are contained in two groups of four atomic
       transitions each. Note that the frequency slope ($\textit{s}$) of four transitions inside the group asymptotically
       tends to the slope of GT. It is easy to show that they are equal to $s = 0.94\un{MHz/G}$ and $s = - 0.94\un{MHz/G}$
       for groups $\textit{1-4}$ and $\textit{5-8}$, respectively.\\
\indent The intensity and the frequency shifts (theory) of the K transitions vs $B$-field for $\pi$-polarized
excitation are shown in Fig.~\ref{fig:14}. As seen for $B \gg B_0$ only transitions $\textit{1-8}$  remain and
the probabilities at $B \gg B_0$ tend asymptotically to the same value as GT (by module $4.66 \times 10^{-18}\un{ESU}$).
The inset shows the diagram of the K $D_1$ line transitions in HPB regime with GT and IFFA marked.
\section{Summary and conclusions}
It is remarkable that with K atoms (due to the small value of $B_0$ for $^{39}$K) contained in NC all the peculiarities of the
behavior of atomic transitions of the Cs, $^{87}$Rb, and $^{85}$Rb in the whole region of $B$-field up to the HPB regime
could be easily studied using smaller, at least by an order, $B$-fields. For example, in Ref.~\cite{Sark4} the $^{87}$Rb $D_1$
line for  $\pi$-polarized excitation was studied (the diagram is similar to the one shown in Fig.~\ref{fig:11}(a)).
For a $B$-field of $220\un{G}$ there was no any evidence of IFFA transitions, while for the K atoms IFFA transitions
(shown in Fig.~\ref{fig:12}) have amplitudes comparable with the other transitions. Also, the peculiarities of the HPB
regime for the K atoms can be detected at $B$-fields smaller by at least one order.
Note that in~\cite{Sarg_2008} a magnetometer based on the Rb NC use was described, while the spatial resolution
of K NC for $L= \lambda/2= 385\un{nm}$ should be better than that of the Rb ($398\un{nm}$) and Cs ($448\un{nm}$),  which is important
when measuring a strongly inhomogeneous magnetic field. It is worth to note that in some cases instead of Rb NC a micrometric-thin ($\sim40\un{\mu m}$)
cell could be successfully used~\cite{Sarg5}.
 Thus, we suppose that with the help of a $40\un{\mu m}$-thin K cell the above mentioned peculiarities could be studied, too.

\acknowledgments
The authors thank A. Papoyan, C. Adams, I. Hughes and H. R. Jauslin for discussions.
The research was conducted in the scope of the International Associated Laboratory IRMAS (CNRS-France $\&$ SCS-Armenia).
A.S. and D.S note that this work was also supported by State Committee Science MESRA, in frame of the research project No. SCS 13-1C029.
A.S. and A.T. thank the ANSEF Fund - Opt 3700 grant.

\end{document}